\documentclass[12pt]{iopart}

\usepackage{graphicx}
\begin{document}

\title[Phase transitions in CeFeAsO, PrFeAsO, and NdFeAsO] {Influence of the rare-earth element on the effects of the structural and magnetic phase transitions in CeFeAsO, PrFeAsO, and NdFeAsO}

\author{Michael A. McGuire$^{1*}$, Rapha\"{e}l P. Hermann$^{2,3}$, Athena S. Sefat$^1$, Brian C. Sales$^1$, Rongying Jin$^1$, David Mandrus$^1$, Fernande Grandjean$^3$, Gary J. Long$^4$}
\address{$^1$Materials Science and Technology Division, Oak Ridge National Laboratory, Oak Ridge, Tennessee 37831 USA}
\address{$^2$Institut f\"ur Festk\"orperforschung, Forschungszentrum J\"ulich GmbH, D-52425 J\"ulich, Germany}
\address{$^3$Department of Physics, B5, Universit\'e de Li\`ege, B-4000 Sart-Tilman, Belgium}
\address{$^4$ Department of Chemistry, Missouri University of Science and Technology, Rolla, MO 65409-0010, USA}
\ead{$^*$McGuireMA@ORNL.gov}

\date{\today}

\begin{abstract}
We present results of transport and magnetic properties and heat capacity measurements on polycrystalline CeFeAsO, PrFeAsO, and NdFeAsO. These materials undergo structural phase transitions, spin density wave-like magnetic ordering of small moments on iron, and antiferromagnetic ordering of rare earth moments. The temperature dependence of the electrical resistivity, Seebeck coefficient, thermal conductivity, Hall coefficient, and magnetoresistance are reported. The magnetic behavior of the materials have been investigated using M\"{o}ssbauer spectroscopy and magnetization measurements. Transport and magnetic properties are affected strongly by the structural and magnetic transitions, suggesting significant changes in the band structure and/or carrier mobilities occur, and phonon$-$phonon scattering is reduced upon transformation to the low temperature structure. Results are compared to recent reports for LaFeAsO, and systematic variations in properties as the identity of Ln is changed are observed and discussed. As Ln progresses across the rare-earth series from La to Nd, an increase in the hole contributions to Seebeck coefficient, and increases in magnetoresistance and the Hall coefficient are observed in the low temperature phase. Analysis of hyperfine fields at the iron nuclei determined from M\"{o}ssbauer spectra indicates that the moment on Fe in the orthorhombic phase is nearly independent of the identity of Ln, in apparent contrast to reports of powder neutron diffraction refinements.
\end{abstract}

\maketitle

\section{Introduction}

Extensive research efforts have been devoted recently to the rare earth iron oxyarsenides. Members of this family have been known for some time \cite{Jeitschko}; however, interest in these and related compounds was renewed by the report of superconductivity at 26 K in doped LaFeAsO \cite{Kamihara}. Isostructural compounds referred to here as LnFeAsO are known to form for Ln = La$-$Gd \cite{Jeitschko}, Tb and Dy \cite{Bos-TbDy}, and also Y \cite{Chong-Y}. Many of the corresponding oxyphosphides and a few oxyantimonides have also been reported, as well as analogues containing formally divalent transition metals other than iron \cite{Jeitschko}. Some oxyphosphides have also been made superconducting \cite{Kamihara2006, Watanabe2007, CavaLaFePO}.

\begin{figure}
\centering
\includegraphics{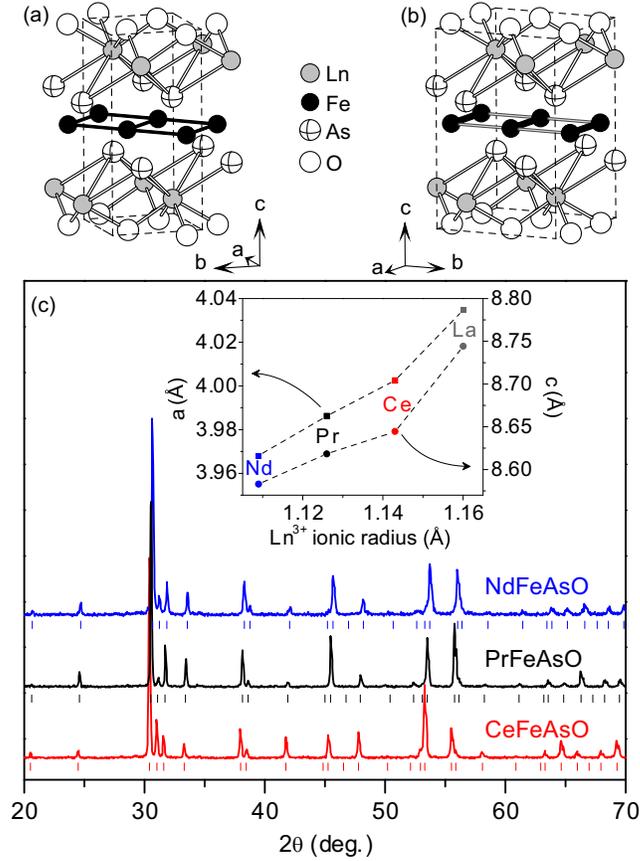}
\caption{\label{fig:structure-PXRD}
Crystal structure of LnFeAsO, drawn to emphasize the distortion of the Fe net at the structural phase transition: (a) high temperature tetragonal phase (space group \textit{P4/nmm}), (b) low temperature orthorhombic phase (\textit{Cmme}). The thicker, darker lines between Fe atoms in (b) indicate shortened Fe$-$Fe contacts. Dotted lines outline the unit cells. (c) Room temperature powder X-ray diffraction patterns from the CeFeAsO, PrFeAsO, and NdFeAsO samples used for the transport and magnetization measurements reported in this work. Tick marks locate reflections belonging to each phase. The inset in (c) shows the refined values of the lattice constants, along with results from LaFeAsO for comparison.
}
\end{figure}

At room temperature these materials adopt the ZrCuSiAs structure type (tetragonal, space group \textit{P4/nmm}), and their structure can be visualized as a stacking of flat square nets of single atom types in the sequence 2O$-$Ln$-$As$-$2Fe$-$As$-$Ln. The Fe and O nets are rotated 45 degrees with respect to the As and Ln nets, and are twice as dense. The stacking of these layers gives tetrahedral coordination to Fe and O, while Ln and As are in square antiprismatic coordination (figure \ref{fig:structure-PXRD}a). The bonding in LnFeAsO is three dimensional$-$there are no van der Waals gaps between layers. However, the states near the Fermi energy have almost pure Fe character, with little mixing of As orbitals, suggesting that these materials can be more two dimensional in their electronic behavior \cite{SinghandDu}.

Superconductivity has been induced in LnFeAsO by doping with fluorine, cobalt \cite{Sefat-Co}, thorium \cite{Th-doping}, strontium \cite{Kasperkiewicz-Nd}, and oxygen vacancies \cite{O-vac-doping}. Maximum reported transition temperatures \textit{T$_c$} at ambient pressure for fluorine doping are 28 K for LaFeAsO$_{0.89}$F$_{0.11}$ \cite{Sefat}, 41 K for CeFeAsO$_{0.84}$F$_{0.16}$ \cite{Chen-Ce}, 52 K for PrFeAsO$_{0.89}$F$_{0.11}$ \cite{Ren-Pr}, 52 K for NdFeAsO$_{0.89}$F$_{0.11}$ \cite{Ren-Nd}, 55 K for SmFeAsO$_{0.9}$F$_{0.1}$ \cite{Ren-Sm}, 36 K for GdFeAsO$_{0.83}$F$_{0.17}$ \cite{Chang-Gd}, 46 K for TbFeAsO$_{0.9}$F$_{0.1}$ \cite{Bos-TbDy}, and 45 K for DyFeAsO$_{0.9}$F$_{0.1}$ \cite{Bos-TbDy}. These compositions are only nominal, based on the loading of the reactants, and in some cases these may not represent optimally doped materials. However, a systematic trend in \textit{T$_c$} is clear, increasing across the rare earth series up to Sm. Thus, understanding the role that Ln plays in determining the physical properties of LnFeAsO based materials is critical.

Soon after the discovery of superconductivity in doped LnFeAsO, studies of the undoped materials revealed interesting structural and magnetic properties. All of the compounds undergo crystallographic transitions from the tetragonal high temperature structure to an orthorhombic low temperature structure at about 150 K, involving a slight distortion of the square nets. In the low temperature phase, magnetic ordering of small moments on the iron sites occurs. When these materials are doped, the structural and magnetic transitions are suppressed and superconductivity emerges. These observations suggest a competition between magnetic and superconducting ground states, and the importance of magnetism to the superconductivity in these compounds \cite{delaCruz-La, McGuire, Zhao-Ce, Zhao-Pr, Qiu-Nd, Chen-Nd}. We have recently reported results from a detailed study of LaFeAsO \cite{McGuire}. The present work focusses on LnFeAsO compounds with Ln = Ce, Pr, and Nd, the first three magnetic rare earth ions. A number of publications and preprints report studies of the crystallographic and magnetic phase transitions in these three materials, primarily utilizing powder neutron diffraction analysis. The results are summarized below.

The structural phase transition is reported to occur near 158 K in CeFeAsO \cite{Zhao-Ce}, 153 K in PrFeAsO \cite{Zhao-Pr}, and 150 K in NdFeAsO \cite{Qiu-Nd}. This transition results in a stretching of the square Fe nets into rectangular nets, and the symmetry is reduced to orthorhombic in the low temperature phase (figure \ref{fig:structure-PXRD}b). Subsequently stripe-like magnetic order of small moments on the Fe sites is observed by neutron diffraction, suggesting the formation of a commensurate spin density wave (SDW). This has been reported at 140 K in CeFeAsO \cite{Zhao-Ce}, 127 K in PrFeAsO \cite{Zhao-Pr}, and 141 K in NdFeAsO \cite{Chen-Nd} Muon spin relaxation measurements yielded 135 K for NdFeAsO \cite{muSR-Nd}. The ordered Fe moment determined from neutron diffraction measurements is small: 0.8(1) $\mu_B$ for CeFeAsO \cite{Zhao-Ce}, 0.48(9) $\mu_B$ for PrFeAsO \cite{Zhao-Pr}, and 0.25(7) $\mu_B$ for NdFeAsO \cite{Chen-Nd}. Several theoretical models involving strong correlations have been proposed to explain the smallness of the Fe magnetic moments. These include consideration of magnetic frustration in both a localized picture \cite{Si} and in \textit{ab initio} density functional theory calculations \cite{Yildirim}, as well as the coupling between a primarily itinerant band and a more localized band \cite{Castro-Neto}. The structural change may be driven by the magnetic interactions; the 1D chains of shorter Fe$-$Fe contacts [figure \ref{fig:structure-PXRD}(b)] allow the stripe$-$like antiferromagnetic ordering to develop. It has been proposed that the structural transition is driven by the removal of magnetic frustration through a distortion of the lattice \cite{Yildirim}, and the existence of a nematic phase in the narrow temperature region between the structural and magnetic transitions has been proposed \cite{nematic}. There is indeed evidence for fluctuating magnetism at high temperatures in some related materials \cite{Norman-Ce}, and this may be responsible for the observed decrease in the magnetization of LaFeAsO upon cooling in the tetragonal phase \cite{McGuire}. Finally, upon further cooling, the rare earth moments in LnFeAsO order antiferromagnetically. This occurs at 4 K in CeFeAsO with the Ce moments lying nearly in the \textit{ab}-plane \cite{Chen-Ce, Zhao-Ce}, at 14 K in PrFeAsO with the Pr moments along the \textit{c}-axis \cite{Zhao-Pr}, and at 2 K in NdFeAsO with the Nd moments canted out of the \textit{ab}-plane \cite{Qiu-Nd}.

There are only a few reports of the transport properties of these compounds, and to our knowledge the reported measurements are limited to electrical resistivity $\rho$ in the absence of an applied magnetic field. All show a cusp in $\rho$ near the structural transition temperature, nearly temperature independent behavior above this temperature, and a rapid decrease below \cite{Kasperkiewicz-Nd, Chen-Ce, Qiu-Nd, Kimber-Pr}. Surprisingly, magnetization measurements have been reported only for NdFeAsO \cite{Kasperkiewicz-Nd}), and the only reported M\"{o}ssbauer spectral study of these compounds is on impure samples of CeFeAsO and PrFeAsO at room temperature \cite{Raffius}.

Here we report a systematic study of the transport and magnetic properties of CeFeAsO, PrFeAsO, and NdFeAsO. Results include the temperature dependence of the electrical resistivity, thermal conductivity, Seebeck coefficient, Hall coefficient, magnetoresistance, and magnetization, as well as results of temperature dependent M\"{o}ssbauer spectral measurements. The results are discussed in the context of the three phase transitions occurring in these materials, and are compared to previously reported results for LaFeAsO.

\section{Experimental Details}
\subsection{Sample synthesis}

NdFeAsO was made from FeAs, Nd$_2$O$_3$, and Nd powders. The rare earth oxide was used as the oxygen source because we have found this route to give the highest purity samples of LaFeAsO \cite{McGuire}. A similar route was attempted for CeFeAsO but was less successful, likely due to the stability of CeO$_2$. Single phase samples of CeFeAsO were made from CeAs, Fe$_2$O$_3$, and Fe powders. Similarly, PrFeAsO was made from PrAs, Fe$_2$O$_3$ and Fe powders. Rare earth arsenides and FeAs starting materials were prepared as reported previously \cite{Sefat, McGuire}. Stoichiometric mixtures of the starting materials (1$-$4 g total masses) were ground and mixed thoroughly in a helium filled glovebox, pressed into one half inch diameter pellets, and sealed in silica tubes under about 0.2$-$0.3 atm ultra-high purity argon. The pellets were heated at 1200 $^\circ$C for 30 h. Powder X-ray diffraction showed the products to be single phase LnFeAsO (limiting impurity concentrations to less than a few percent) with the ZrCuSiAs structure type (figure \ref{fig:structure-PXRD}c).

\subsection{Characterization techniques}

Powder X-ray diffraction was performed on a Scintag 2000 diffractometer using Cu K$\alpha$ radiation, and lattice constants were determined from full pattern LeBail refinements using the program FullProf \cite{FullProf}. Transport measurements were performed using a Physical Property Measurement System from Quantum Design. The samples were contacted with silver epoxy (Epotek H20E). Gold coated copper leads were used for Seebeck coefficient and thermal conductivity measurements. Platinum leads were used for resistivity, Hall effect, and magnetoresistance measurements. Hall coefficients were determined from transverse resistivity measurements in applied fields from -6 T to 6 T. Magnetization measurements were performed using a SQUID magnetometer (Quantum Design, Magnetic Properties Measurement System). The iron-57 M\"{o}ssbauer spectra of LnFeAsO were recorded between 4.2 (2.9 for Ln = Ce) and 295 K on a constant acceleration spectrometer equipped with a rhodium matrix cobalt-57 source and calibrated with $\alpha$-iron powder at 295 K, which was also used as an isomer shift reference.

\section{Results and discussion}

Powder X-ray diffraction patterns collected at room temperature for the LnFeAsO samples are shown in figure \ref{fig:structure-PXRD}c. These same samples were used for all of the measurements reported below. Tick marks show predicted peak positions for each phase in the space group \textit{P4/nmm} with refined lattice constants (\textit{a} and \textit{c}) of 4.0025(3) {\AA} and 8.6441(4) {\AA} for CeFeAsO, 3.9862(2) {\AA} and 8.6184(5) {\AA} for PrFeAsO, and 3.9681(4) {\AA} and 8.5830(9) {\AA} for NdFeAsO, in reasonable agreement with reported values \cite{Jeitschko}. For comparison, the room temperature lattice constants of LaFeAsO are a = 4.03463(5) {\AA} and c = 8.74382(15) {\AA} \cite{McGuire}. The inset of figure \ref{fig:structure-PXRD}c shows the variation of the lattice constants with the ionic radius of the corresponding trivalent rare earth ion in eight-fold coordination \cite{Shannon}. The deviation from a smooth trend at Ln = Ce originally observed by Quebe \textit{et al.} \cite {Jeitschko} is noted. This may indicate the presence of some Ce$^{4+}$ (a few percent) which has a smaller ionic radius of 0.97 {\AA} \cite{Shannon}. Heavy fermion behavior has been observed in the related compound CeFePO, with strong electron correlations arising from Ce-4\textit{f} electrons \cite{CeFePO}.

\begin{figure}
\centering
\includegraphics[width=4.0in]{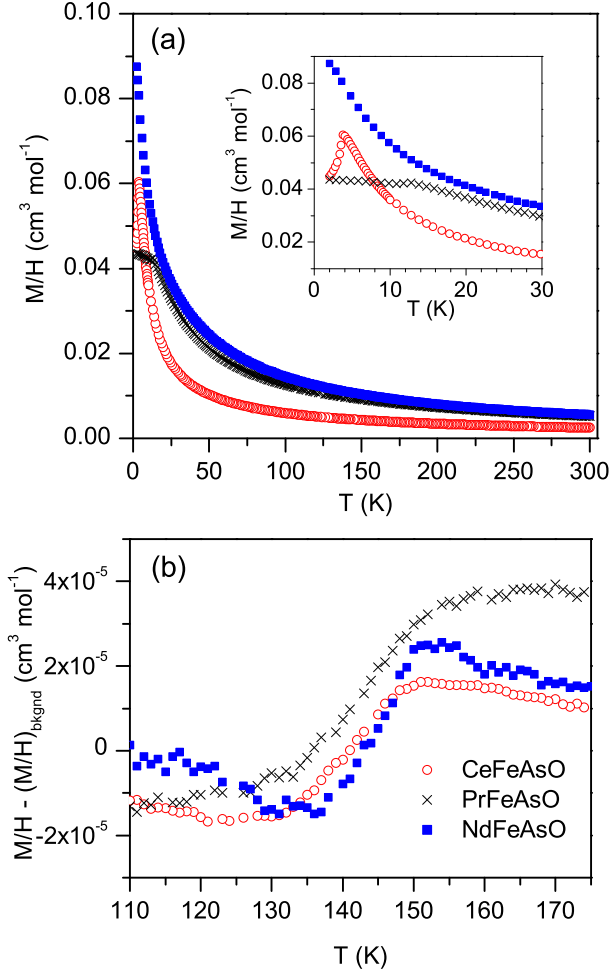}
\caption{\label{fig:Mag}
Magnetic properties of CeFeAsO, PrFeAsO, and NdFeAsO. (a) Temperature dependence of \textit{M/H} at an applied field of 1 T. The inset in (a) shows the behavior of \textit{M/H} near the rare earth ordering temperatures. (b) \textit{M/H} near the structural phase transitions and the Fe magnetic ordering temperatures after subtracting a Curie-Weiss-type ``background'' \textit{(M/H)$_{bkgnd}$} fit through the transition region. Note that the negative values in (c) do not indicate a negative magnetization, only a negative deviation from the smooth background fit.
}
\end{figure}

The measured magnetic behaviors of CeFeAsO, PrFeAsO, and NdFeAsO are summarized in figure \ref{fig:Mag}. Figure \ref{fig:Mag}a shows the temperature dependence of M/H for all three materials, where M is the magnetization and H is the applied field. The magnetization of these materials is dominated by the Ln moments. Evidence of antiferromagnetic ordering of Ce and Pr at T$_N$ = 3.8 K and 13 K, respectively, is emphasized in the inset of figure \ref{fig:Mag}a. These values agree with the reported ordering temperatures determined by neutron diffraction and heat capacity measurements \cite{Zhao-Ce, Chen-Ce, Zhao-Pr}. Above the rare earth and iron ordering temepratures, the data are well described by a Curie Weiss law with an additional temperature independent term. Fits to the data using this model from 200 to 300 K give effective moments and Weiss temperatures of 2.25(4) $\mu_B$ and -16(2) K for CeFeAsO, 3.64(2) $\mu_B$ and -36(1) K for PrFeAsO, and 3.60(1) $\mu_B$ and -10(1) K for NdFeAsO. Although the polycrystalline nature of these materials and possible low lying crystal field levels complicate the analysis of the effective moments derived from these measurements, nice agreement is observed between the fitted moments and the free ion values of 2.54 $\mu_B$ for Ce, 3.58 $\mu_B$ for Pr, and 3.62 $\mu_B$ for Nd. In addition, the fitted Weiss temperatures scale with the observed Ln ordering temperatures, as expected.

In an effort to observe the effects of the Fe moments on the magnetization of these materials, the deviations between the data and a smooth background generated from a Curie-Weiss fit through the transition region were examined. As shown in figure \ref{fig:Mag}b, a sharp drop in this ``residual'' magnetization is observed in each material near 150 K. The decrease is of comparable magnitude in each compound (3$-$4 $\times 10^{-5} cm^3 mol^{-1}$). Very similar behavior is observed in LaFeAsO \cite{McGuire}, in which this decrease in \textit{M/H} is estimated to be 4 $\times 10^{-5} cm^3 mol^{-1}$. The reported structural transition temperatures range from 150 to 160 K in these materials, and the magnetic transition temperatures range from about 130 to 140 K. The onsets of the observed magnetic anomalies (figure \ref{fig:Mag}b) are near 150 K and their centers are near 140$-$145 K. The closely spaced structural and magnetic transition temperatures, along with the gradual nature of the crystallographic transition reported in some of these materials \cite{McGuire, Fratini-Nd}, makes it difficult to assign the drop in \textit{M/H} to one of the transitions. However, the freezing of fluctuating magnetism into the SDW state may be responsible for the observed behavior.

\begin{figure}
\centering
\includegraphics[width=2.5in]{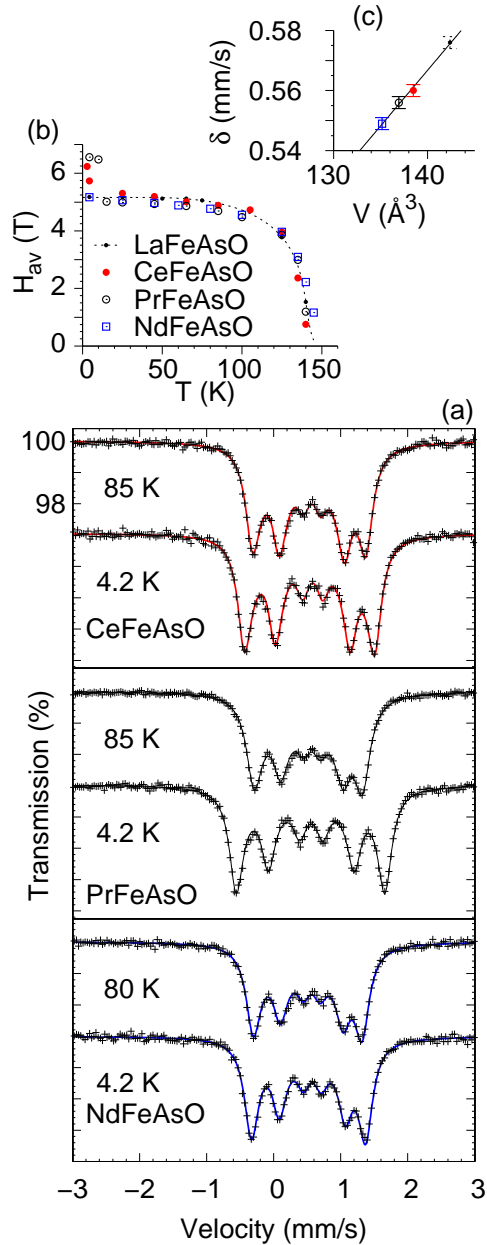}
\caption{\label{fig:Moss}
(a) The Mössbauer spectra of LnFeAsO in the orthorhombic phase, (b) The temperature dependence of the average hyperfine field, (c) The dependence of the isomer shift at 4.2 K upon rare-earth substitution as a function of the unit-cell volume.
}
\end{figure}

The M\"{o}ssbauer spectra of LnFeAsO below the Fe magnetic ordering temperature are shown in figure \ref{fig:Moss}a, and are similar to reported spectra from LaFeAsO \cite{McGuire, AA, Raffius, Kitao}. Analysis of the spectra reveal no iron bearing secondary phases in these materials at a detection limit of ca. 1\%. This technique has been shown to be very useful in identifying iron containing impurities in this class of compounds \cite{Nowik}, and is expected to be more sensitive to such secondary phases than PXRD using Cu K$\alpha$ radiation. The spectra in figure \ref{fig:Moss}a have been fit with a hyperfine field distribution (details will be reported in a separate publication \cite{Raphael}). The temperature dependence of the average hyperfine field observed up to 150 K is given in Fig. \ref{fig:Moss}(b). The isomer shift at 4.2 K exhibits a significant increase with increasing Ln ionic radius, and thus increasing unit-cell volume, see Fig. \ref{fig:structure-PXRD}(c) and Fig. \ref{fig:Moss}(c). This increase in isomer shift is correlated with the decrease of the \textit{s}-electron density at the Fe nucleus. The quadrupole shift is, on average, -0.04(1), -0.06(1), and -0.06(1) mm/s for Ln = Ce, Pr, and Nd, respectively; shifts slightly more negative than is observed for LaFeAsO \cite{McGuire}. The combination of isomer shift values, small absolute values of the quadrupole shift, and small hyperfine fields are most similar to a low spin iron(II) electronic configuration, although this ionic picture is likely not appropriate for these compounds.

At the lowest temperatures investigated here, an increase in the hyperfine field is observed for CeFeAsO and PrFeAsO, upon ordering of the rare-earth moment [Fig. \ref{fig:Moss}(b)]. This increase likely corresponds to a small additional rare-earth transferred field. From the M\"{o}ssbauer spectral data, we estimate the rare-earth ordering temperature to be near 5 and 12 K for CeFeAsO and PrFeAsO, respectively, in agreement with previous reports and with the magnetization results shown in Fig. \ref{fig:Mag}(a). Upon ordering of Ce, the average hyperfine field at iron increases by 0.9 T and no significant change in the quadrupole shift is observed. In contrast, upon Pr ordering, the average hyperfine field increases by 1.6 T and the quadrupole shift increases significantly  from -0.06(1) mm/s at 15 K to -0.007(6)  mm/s at 4.2 K. This increase in quadrupole shift is indicative of a spin reorientation, that results in a reorientation of the iron hyperfine field in the principal axes of the electric field gradient, a phenomenon frequently encountered in Ln-Fe compounds with competing magnetic sublattices. This reorientation is easily observed by M\"{o}ssbauer spectroscopy \cite{AA, BB}. This observation is consistent with the reports that the ordered Ce moments are nearly in the \textit{ab}-plane \cite{Chen-Ce, Zhao-Ce} and the Pr moments are along the \textit{c}-axis \cite{Zhao-Pr}. The observed spin reorientation indicates that it is very likely that the Fe moments in PrFeAsO have a significant component along the c-axis below the Pr ordering temperature.

We note also that the average hyperfine field, \textit{H$_{av}$}, 5.17(1), 5.06(2), 4.99(1), and 5.30(1) T at 25 K for Ln = La, Ce, Pr, and Nd, respectively, observed above the rare-earth ordering temperatures are only weakly dependent on Ln. This indicates that there is little change in the ordered magnetic moment on Fe in these materials as Ln is varied. Using the usual conversion factor of 15 T per 1 $\mu_B$ gives estimated Fe local moments ranging between 0.33 and 0.35 $\mu_B$. The ordered magnetic moment on Fe determined by neutron diffraction for LaFeAsO [0.36(5) $\mu_B$] \cite{delaCruz-La} and NdFeAsO [0.25(7) $\mu_B$] \cite{Qiu-Nd} agrees reasonably well with this local moment inferred from the M\"ossbauer spectra. Interestingly, significant disagreement is observed between the Fe moment determined from neutron diffraction for CeFeAsO [0.8(1) $\mu_B$] \cite{Zhao-Ce} and PrFeAsO [0.48(9) $\mu_B$] \cite{Zhao-Pr} and the behavior observed here. The source of these discrepancies is unclear. We note, however, that the results presented here are consistent with previous reports from $\mu$SR measurements which also deduced similar ordered Fe magnetic moments in LaFeAsO and NdFeAsO \cite{muSR-Nd}.

\begin{figure}
\centering
\includegraphics[width=3.5in]{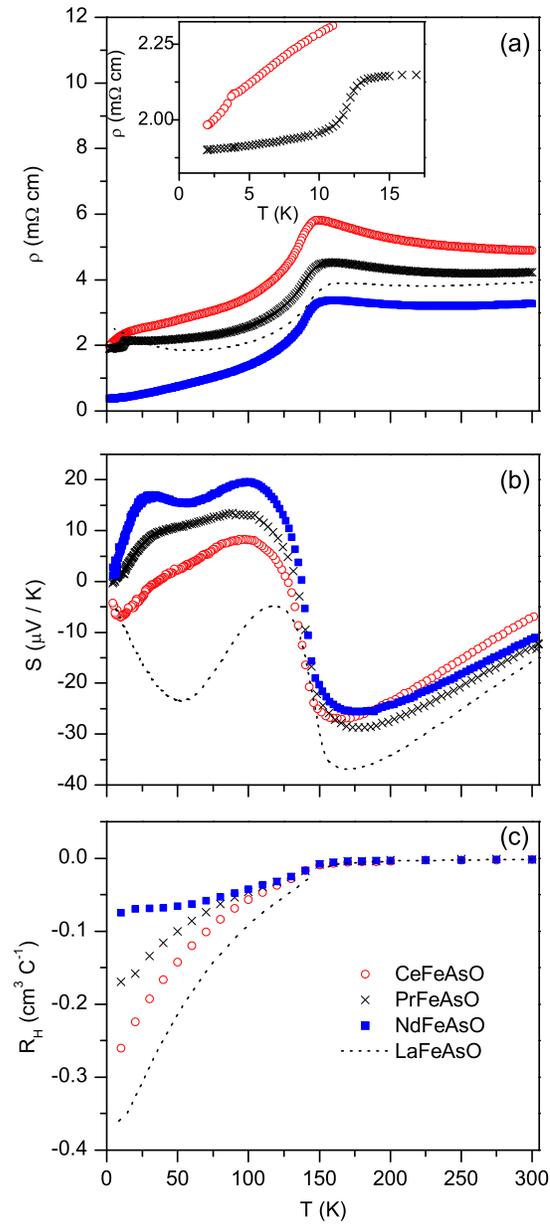}
\caption{\label{fig:transport}
The temperature dependence of the transport properties of CeFeAsO, PrFeAsO, and NdFeAsO. Data for LaFeAsO are shown for comparison. (a) Electrcial resistivity $\rho$, with the behavior near the Ce and Pr ordering temperatures shown in the inset. (b) Seebeck coefficient \textit{S}. (c) The Hall coefficient \textit{R$_H$}.
}
\end{figure}

\begin{figure}
\centering
\includegraphics[width=4.0in]{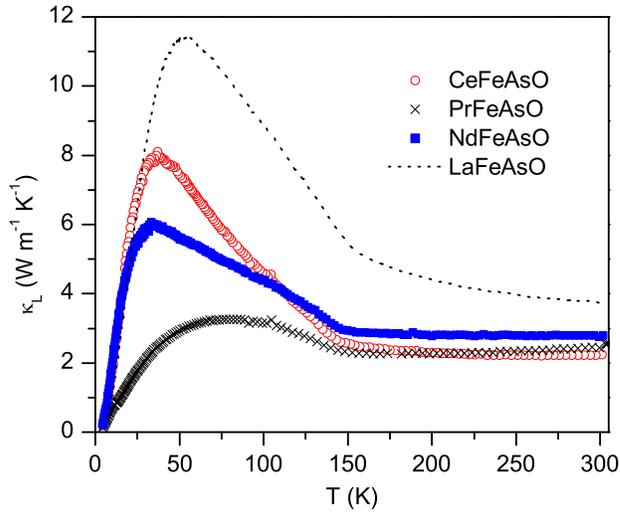}
\caption{\label{fig:kappa}
The lattice thermal conductivity $\kappa_L$ of CeFeAsO, PrFeAsO, and NdFeAsO. Results for LaFeAsO are included for comparison.
}
\end{figure}

The results of transport properties measurements on CeFeAsO, PrFeAsO, and NdFeAsO are shown in figures \ref{fig:transport} and \ref{fig:kappa}. Data from Ref. \cite{McGuire} for LaFeAsO are shown for comparison. The polycrystalline nature of these samples complicates the comparison of the actual values of the resistivity and thermal conductivity. Qualitatively similar behavior is observed in all of these materials; however, important trends are observed as Ln is varied.

The resistivity (figure \ref{fig:transport}a) is nearly temperature independent in the tetragonal phase. At about 150 K all of the materials show a maximum in $\rho$, and upon further cooling a relatively sharp drop in $\rho$ occurs. Additional downturns are observed at low temperatures for PrFeAsO and CeFeAsO (figure \ref{fig:transport}a, inset), reflecting decreased carrier scattering as the Ln magnetic moments order. Data were collected only at temperatures above 2 K, so effects of the Nd ordering are not observed.

The Seebeck coefficient S (figure \ref{fig:transport}b) shows interesting and systematic behavior as Ln is varied. There is a sharp increase in \textit{S} as the materials are cooled below their structural phase transition temperatures. These are multiband materials, with several hole and electron pockets contributing to the electrical transport \cite{SinghandDu}. The Seebeck coefficient is then a sum of each band's contribution, weighted by the conductivity of the individual bands. With this in mind, the rapid increase in \textit{S} may indicate a dramatic change in the charge carrier scattering mechanism, or a reduction in the contribution of an electron band to the total \textit{S}, or similarly an increase in the contribution from a hole band. Unlike LaFeAsO, the Seebeck coefficient of the Ce, Pr, and Nd compounds actually becomes positive below 150 K, and remains positive down to 4 K for PrFeAsO and NdFeAsO. This suggests a transition from electron dominated conduction to hole dominated conduction as the band structure responds to tuning of the unit cell dimensions. A low temperature peak in \textit{S} at about 30 K can be seen ``growing in'' as Ln progresses from across the series. The local maximum in S for NdFeAsO near 30 K and similar features in the CeFeAsO and PrFeAsO data are reminiscent of the low temperature behavior of some Ce and Yb intermetallic compounds, and may be related to crystal field splitting of the 4\textit{f} ground state multiplet \cite{Zlatic}.

The Hall coefficient \textit{R$_H$}, shown in figure \ref{fig:transport}c is negative in all of these materials over the entire temperature range studied. R$_H$ is only weakly temperature dependent in the tetragonal phase, while in the orhtorhombic phase (below about 150 K) \textit{R$_H$} increases rapidly in magnitude upon cooling. The temperature dependence of $R_H$ will be further discussed below. The identity of Ln dramatically affects the behavior of \textit{R$_H$}. Interpretation of Hall coefficient data is complicated by the presence of multiple bands at the Fermi level. Carrier concentrations and relative mobilities of the bands affect the value of the Hall coefficient. In the simplest model with one electron band and one hole band with equal mobilities, \textit{R$_H$} is expected to increase with increasing hole concentration for a fixed electron concentration. Within this simple approximation the dependence of \textit{R$_H$} on the identity of Ln shown in figure \ref{fig:transport}c is consistent with the Seebeck coefficient results discussed above (figure \ref{fig:transport}b). The observed influence of the rare earth element on \textit{R$_H$} is further evidence of the significant effects of Ln on the electronic properties of the Fe layers in LnFeAsO materials, primarily through contraction of the lattice as the ionic radius of Ln decreases.

The lattice thermal conductivity $\kappa_L$ of the LnFeAsO materials are shown in figure \ref{fig:kappa}. The electronic contribution was subtracted from the measured total thermal conductivity using the Wiedemann-Franz law. In general the data show behavior typical of crystalline materials, an increase with temperature at low temperatures followed by a decrease with temperature as Umklapp scattering begins to dominate the phonon$-$phonon interactions. Differences in the height of the low temperature peak in $\kappa_L$ may simply reflect the polycrystalline nature of the materials. Superimposed upon this ``normal'' behavior is a change in slope near 150 K; $\kappa_L$ increases abruptly below the structural phase transition temperature. This suggests that when the structure changes from tetragonal to orthorhombic the phonon scattering rate is decreased. This could be due to the freezing out of structural fluctuations or soft phonon modes related to the impending crystallographic phase change.

\begin{figure}
\centering
\includegraphics[width=4.0in]{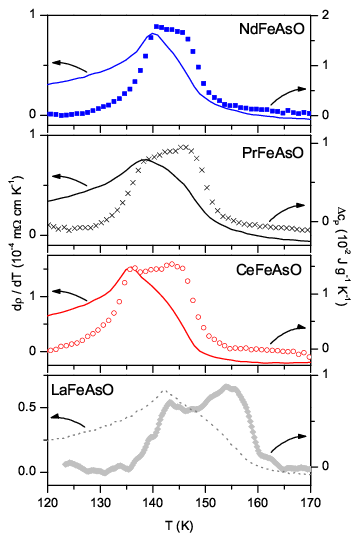}
\caption{\label{fig:drdT-HC}
Temperature derivative of the resistivity (d$\rho$/dT) and the background-subtracted heat capacity ($\Delta$c$_P$). Peaks corresponding to the structural transitions and Fe magnetic ordering transitions are observed in both of these properties.
}
\end{figure}

Due to the smooth variation of the transport properties and the magnetic and structural transition temperatures it is difficult to determine whether the observed anomalies are related to the structural transition, the magnetic transition, or both. However, inspection of the temperature derivative of the electrical resistivity shown in figure \ref{fig:drdT-HC} reveals that effects of both the structural and magnetic transitions can be observed. Two overlapping peaks are identified, one due to the crystallographic distortion, and one due to onset of long range magnetic ordering. Also shown in this figure is the background subtracted specific heat data. A polynomial fit to data above and below the transition region was subtracted from the measured data to emphasize the effects of the phase transitions. Two overlapping peaks in the specific heat are clearly visible in figure \ref{fig:drdT-HC}, similar to previous reports for LaFeAsO \cite{McGuire, Kohama}. Nice agreement is observed between the temperatures at which these peaks and the peaks in \textit{d$\rho$/dT} occur. This demonstrates that the structural and long range magnetic ordering transitions are indeed distinct in these materials, and that each is responsible for significant changes in the transport properties of LnFeAsO. Taking the cusps in the specific heat as estimates of the transition temperatures shows that the magnetic ordering temperature increases from 136 K for CeFeAsO to 139 K for PrFeAsO to 141 K for NdFeAsO. The structural transition, defined in this way, varies somewhat less strongly across the series from 144 K for CeFeAsO to 146 K for both PrFeAsO and NdFeAsO. It is interesting that these values vary systematically from Ce to Nd, whereas the transition temperatures determined by the same method for LaFeAsO (figure \ref{fig:drdT-HC}d) do not follow this trend. This different behavior between materials containing non-magnetic and magnetic rare-earth elements suggests that the presence of the magnetic moment on the Ln site influences the structural distortion and magnetic ordering.

\begin{figure}
\centering
\includegraphics[width=4.0in]{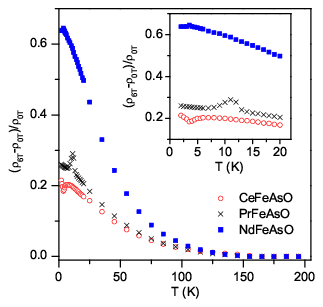}
\caption{\label{fig:MR}
The magnetoresistance of CeFeAsO, PrFeAsO, and NdFeAsO at an applied field of 6 T. The inset shows the behavior of the magnetoresistance near the Ln magnetic ordering temperatures.
}
\end{figure}

Figure \ref{fig:MR} shows the magnetoresistance of CeFeAsO, PrFeAsO, and NdFeAsO, with the applied magnetic field perpendicular to the direction of the electrical current. We note that the polycrystalline nature of the samples may significantly influence the magnetoresistance through variations in grain size and intergrain coupling; however, systematic behavior is observed from across the Ln series Ce$-$Pr$-$Nd. The magnetoresistance increases abruptly below the structural and magnetic phase transitions, due to the influence of the spin density wave on the charge carriers. The maximum magnetoresistance realized at low temperatures increases across the rare earth series Ce$-$Pr$-$Nd. Again, effects on the electronic structure caused by the varying degree of compression of the Fe planes across the Ln series is expected to be important, and may be reflected in the magnetoresistance. Variation of the number of Ln f-electrons and the associated magnetic moments may also influence the observed behavior. Anomalies in the magnetoresistance are observed at the rare earth ordering temperatures (figure \ref{fig:MR} inset), reflecting the field dependence of the N\'{e}el temperatures of the Ce and Pr sublattices.

The results presented above suggest that significant changes in the electronic structure of LnFeAsO occur between the high temperature tetragonal phase and the low temperature orthorhombic phase with SDW like antiferromagnetic order. Two simple scenarios describing the change in the electronic structure upon cooling through the phase transitions will be considered here. The first involves a decrease in the density of states (DOS) at the Fermi energy (E$_F$) due to the removal of an electron pocket which may be expected to accompany SDW formation, the second involves an increase in the DOS at E$_F$ due to the addition a new hole pocket which has been proposed to occur at the structural transition based on some electronic structure calculations \cite{Yildirim}.

We  note that either the removal of an electron pocket, or the addition of a new hole pocket upon cooling through the structural and SDW transitions are consistent with the observed behavior of the Seebeck coefficient. At lower temperatures the behavior is indicative of the presence of competing electron and hole bands, while above the transitions the behavior is dominated by electron bands. Other observed transport behaviors are discussed in the context of these two scenarios below.

Removal of an electron pocket and decrease in the DOS at E$_F$ is, in the simplest single-band approximation, consistent with the Hall coefficient results, which indicate a decrease in the inferred concentration of electrons below about 150 K. Although this is certainly an oversimplification, as multiband effects are clearly important. In this scenario, the decrease in electrical resistivity must be due to an decrease in charge carrier scattering, which overwhelms the effects of the decrease in carrier concentration. This would indicate strong scattering is present in the high temperature phase. Possible sources of strong scattering are lattice fluctuations related to the impending structural transition, also likely responsible for the observed behavior of $\kappa_L$, and electron-electron scattering involving electrons which are gapped out at lower temperature due to the SDW transition.

The appearance of a new hole pocket and increase in the DOS at E$_F$ below the structural/SDW transition neatly explains the decreased resistivity in the low temperature phase, and the behavior of the Seebeck coefficient as discussed above. The decrease in the Hall coefficient with temperature at low temperatures would then be attributed to its complex response in multi-band materials, while the behavior of $\kappa_L$ can be attributed simply to lattice fluctuations in the high temperature phase.

Clearly it is difficult to reconcile the temperature dependence of S and R$_H$ through the structural distortion and Fe magnetic ordering using simple models. This is not unexpected since five bands are likely contributing to the electronic conduction, and the crystallographic and magnetic phase transitions may strongly change carrier scattering rates and mechanisms. Correlations between experimental results and predictions of first principles electronic structure calculations could prove very helpful in understanding the electronic and magnetic nature of LnFeAsO. This could be especially enlightening regarding the trends observed as the identity of Ln is varied. However, at present no consensus has been reached among the numerous theoretical studies of these materials, due in part to remarkable sensitivities to structural and calculational details \cite{Mazin, Lebegue}.

\section{Conclusions}

Our systematic investigation into the effects of the structural and magnetic phase transitions in CeFeAsO, PrFeAsO, and NdFeAsO revealed several interesting behaviors and relationships. Structural and SDW transition temperatures vary little, but systematically, as Ln is varied from Ce to Pr to Nd. However, they are significantly lower than those in LaFeAsO, suggesting a strong influence of the rare earth magnetic moment on the behavior of the FeAs layers. The phase transitions result in dramatic changes in the transport properties of these materials, similar to those observed previously in LaFeAsO. A sharp drop in resistivity and increase in lattice thermal conductivity are observed upon cooling though the structural and SDW transitions. The behavior of the Hall coefficient, Seebeck coefficient, and magnetoresistance vary systematically across the rare-earth series. Seebeck coefficient measurements reveal increasingly hole-dominated conduction at low temperatures as the lattice constants are decreased due to varying the Ln ionic radii. The Hall coefficient results, though difficult to interpret in multi-band materials, are also consistent with increasing hole concentration at low temperatures as Ln varies from La to Nd, at least in the simplest two-band approximation. To date, no real consensus has been reached among theorists concerning the electronic nature of these complex materials and their phase transitions. This is especially true concerning the Fe magnetism in the orthorhombic phase. Regarding the Fe magnetism, we find from an analysis of the M\"{o}ssbauer spectra in the SDW state above the Ln ordering temperatures an average hyperfine field nearly independent of Ln, in apparent contrast to previous reports based on powder neutron diffraction. The M\"{o}ssbauer spectra also reveal interactions between the Ln ions and the Fe layers through an abrupt increase in the hyperfine field at Fe below the Ln magnetic ordering temperatures, and a Fe spin reorientation in the case of PrFeAsO.

\section{Acknowledgements}

We are grateful to M. T. Sougrati for help with acquisition of the M\"{o}ssbauer spectra. Research sponsored by the Division of Materials Sciences and Engineering, Office of Basic Energy Sciences.
Part of this research performed by Eugene P. Wigner Fellows at Oak Ridge National Laboratory, managed by UT-Battelle, LLC, for the U.S. DOE under Contract DE-AC05-00OR22725. Work in Liege is supported by the Fonds National de la Recherche Scientifique, Belgium, through Grants No. 9.456595 and No. 1.5.064.05.

\section*{References}

\bibliography{LnFeAsO}

\end{document}